\documentclass[twocolumn]{aa}  

\usepackage{graphicx}
\usepackage{txfonts}
\usepackage{booktabs}

\usepackage[]{hyperref}
\begin{document} 

    \title{Nested active regions anchor the heliospheric current sheet and stall the reversal of the coronal magnetic field}

   \titlerunning{Nested Active Regions Influence the HCS and Coronal Field Reversal}
   \authorrunning{Finley et al.}

   \author{A. J. Finley 
          }

    \institute{Universit\'e Paris-Saclay, Universit\'e Paris Cit\'e, CEA, CNRS, AIM, 91191, Gif-sur-Yvette, France\\ \email{adam.finley@cea.fr} }

   \date{Received Sept 30, 2024; accepted -- --, 202-}

\abstract{During the solar cycle, the Sun's magnetic field polarity reverses due to the emergence, cancellation, and advection of magnetic flux towards the rotational poles. Flux emergence events occasionally cluster together, although it is unclear if this is due to the underlying solar dynamo or simply by chance.}
{Regardless of the cause, we aim to characterise how the reversal of the Sun's magnetic field and the structure of the solar corona are influenced by nested flux emergence.}
{From the spherical harmonic decomposition of the Sun's photospheric magnetic field, we identify times when the reversal of the dipole component stalls for several solar rotations. Using observations from sunspot cycle 23 to present, we locate the nested active regions responsible for each stalling and explore their impact on the coronal magnetic field using potential field source surface extrapolations.}
{Nested flux emergence has a more significant impact on the topology of the coronal magnetic field than isolated emergences as it produces a coherent (low spherical harmonic order) contribution to the photospheric magnetic field. The heliospheric current sheet, that separates oppositely directed coronal magnetic field, can become anchored above nested active regions due to the formation of strong opposing magnetic fluxes. Further flux emergence, cancellation, differential rotation, and diffusion, then effectively advects the heliospheric current sheet and shifts the dipole axis.}
{Nested flux emergence can restrict the evolution of the heliospheric current sheet and impede the reversal of the coronal magnetic field. The sources of the solar wind can be more consistently identified around nested active regions as the magnetic field topology remains self-similar for multiple solar rotations. This highlights the importance of identifying and tracking nested active regions to guide the remote-sensing observations of modern heliophysics missions.}

   \keywords{Solar Magnetism -- Solar Cycle -- Solar Wind --
                    }

   \maketitle
%

\section{Introduction}

The Sun's magnetic field is maintained by an internal dynamo that modulates solar activity over an approximately 11 year cycle \citep[][]{hathaway2015solar}. The dynamo relies on the shearing of magnetic field by differential rotation and the subsequent rise of new magnetic flux through the convection zone to form active regions; visible as sunspots \citep[see review of][]{brun2017magnetism}. Active regions emerge in latitudinal bands that migrate towards the equator during the solar cycle. The north-south tilt of emerging active regions favours flux cancellation near the equator and the migration of opposite polarity magnetic flux to the rotational poles \citep{schunker2020average}. This weakens and then reverses the Sun's polar fields, causing the dipole component of the Sun's magnetic field to flip during the sunspot cycle \citep{ sun2015polar}. However, at solar maximum, when the polar fields are weak, the magnetic flux from active regions contributes significantly to the Sun's dipole field \citep{finley2023evolution}. Therefore, the path traced by the dipole axis during the solar cycle is highly structured \citep[see Figure 3 of][]{derosa2012solar}.  

Nested active region emergence, where magnetic flux rises up in the vicinity of recently emerged active regions, is frequently observed on the Sun \citep{ribes1993solar, gaizauskas1994interactions, castenmiller1986sunspot, brouwer1990sunspot, pojoga2002clustering, usoskin2005preferred, gaizauskas2008development}. The nesting of active regions has also become relevant for interpreting the brightness variations of Sun-like stars, as nesting creates a strong rotational modulation \citep{sowmya2021predictions, breton2024tracking}. Whether nesting relates to the underlying dynamo or is a stochastic feature of flux emergence is currently debated. However, it is clear that a significant fraction of active regions, perhaps up to 50\%, emerge in this way on the Sun \citep{schrijver2008solar}. The formation of long-lived nested active regions appears to be coherent across sunspot cycles, with active longitudes in each hemisphere separated by 180$^{\circ}$. These longitudes drift relative to the Carrington reference frame due to the Sun's latitudinal differential rotation \citep{berdyugina2003active, tahtinen2024straight}. 

\begin{figure*}
    \centering
    \includegraphics[trim=0cm 0cm 0cm 0cm, clip, width=\textwidth]{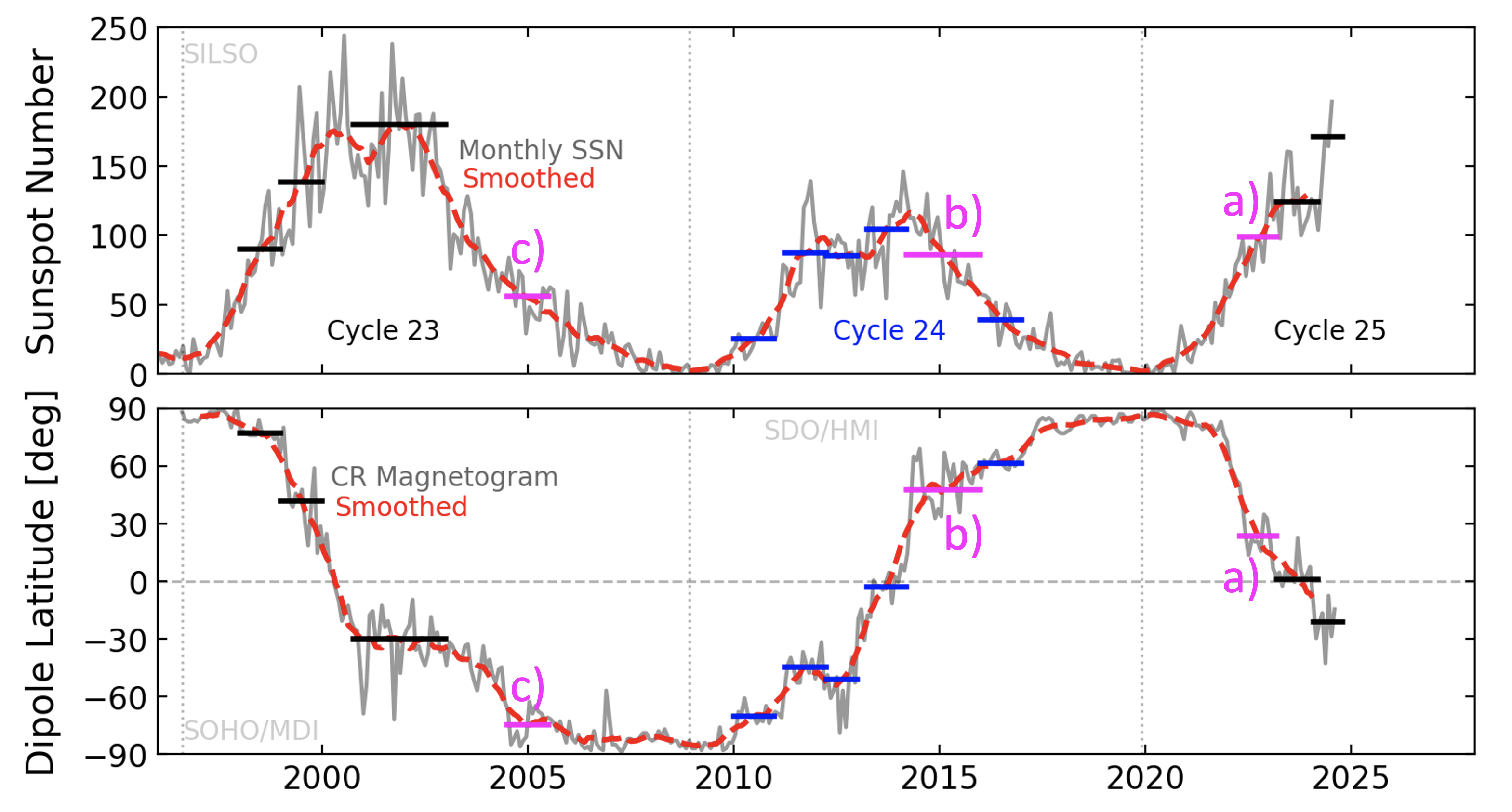}
    \caption{Time-series of sunspot number and positive dipole axis latitude since the beginning of sunspot cycle 23. The grey solid lines have a monthly, or Carrington rotation, cadence. The red dashed lines show the 13-month smoothed version of each time-series. The start time of sunspot cycle 23, 24, and 25 are identified with dotted vertical lines. Horizontal bars indicate when the reversal of the dipole component is halted for several solar rotations. The bars coloured magenta and labelled a), b), and c) are discussed in Section 3.}
    \label{fig:br_lat}
\end{figure*}

\begin{figure*}
    \centering
    \includegraphics[trim=0cm 0cm 0cm 0cm, clip, width=\textwidth]{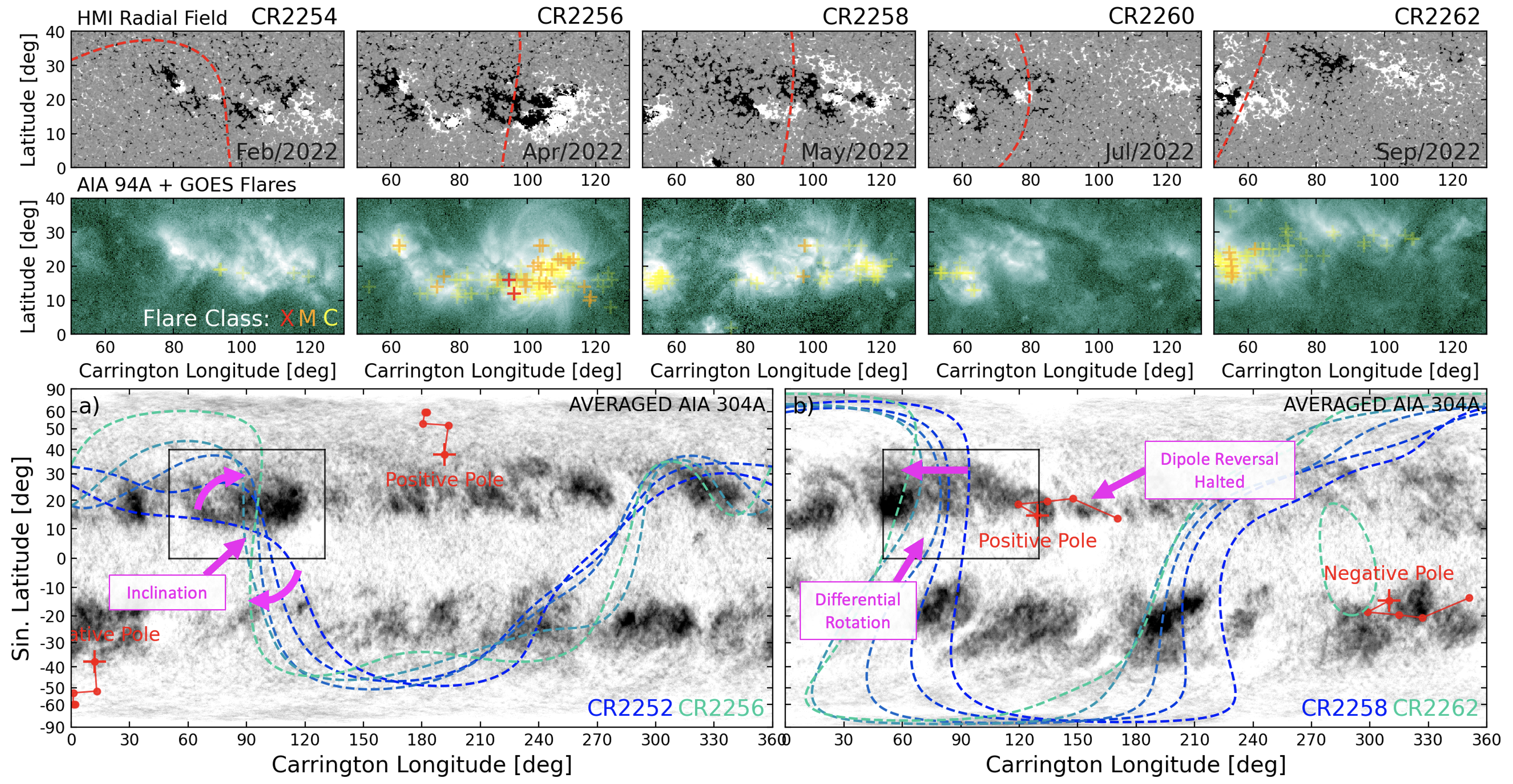}
    \caption{Summary of the magnetic field and activity evolution of a long-lived nested active region spanning January 2022 to October 2022. The top two rows show five snapshots of the HMI radial magnetic field and AIA 94$\AA$ emission, with each snapshot taken two Carrington rotations apart. The dashed red line over the magnetic field shows the location of the HCS at the top of the PFSS extrapolation. Coloured cross markers over the EUV emission show the location of flares detected by GOES, and their associated flare class by colour. The background of the two lower panels display the average 304$\AA$ emission from five consecutive Carrington maps (each spanning one solar rotation), with darker tones having more intense emission. Panel a) spans CR 2252 to 2256 and panel b) spans CR 2258 to 2262. A solid black rectangle shows the selected area from the snapshots above. The evolution of the dipole axis is shown with connected red dots that track the position of the positive and negative poles. The crossed dot indicates the final position in the sequence. Dashed coloured lines follow the evolution of the HCS in time, moving from blue through to green. Panel a) focuses on the emergence phase, and panel b) the subsequent decay and differential rotation of the region.}
    \label{fig:fig_a}
\end{figure*}

Predicting the structure of the coronal magnetic field is important for missions such as NASA's Parker Solar Probe \citep{fox2016solar} and ESA's Solar Orbiter \citep{muller2020solar}, that aim to connect in-situ solar wind measurements from the inner heliosphere to their source regions. Modelling the coronal magnetic field with a potential field source surface (PFSS) model \citep{altschuler1969magnetic, schrijver2003photospheric} and extrapolating the connectivity to the spacecraft, either using an analytic form of the Parker-spiral \citep{badman2020magnetic} or numerically accounting for fast-slow wind-stream interactions \citep{barnard2022huxt}, has become the standard operational technique for these missions \citep{rouillard2020models}. More advanced coronal models are available that solve the full magnetohydrodynamic equations \citep{riley2019predicting, reville2020role, perri2022coconut}, and couple to heliospheric models \citep{riley2011global, wold2018verification, gombosi2018extended, verbeke2022icarus, baratashvili2024operational}. However, such models are computationally expensive and struggle to explore the full range of possible input photospheric magnetic field configurations. Active regions emerging out of view on the Sun's far-side also introduce significant challenges for modelling the corona and solar wind \citep{perri2024impact}. Although, Solar Orbiter now periodically acts as far-side monitor and ESA's Vigil mission (originally known as Lagrange) is set to observe the Sun from the fifth Sun-Earth Lagrange point in the 2030s. Vigil will provide an early warning of magnetic features rotating into view of Earth from the east-limb.

The accuracy of spacecraft connectivity is often limited by uncertainty in the location of the heliospheric current sheet (HCS). The HCS divides oppositely directed coronal magnetic field that originate from very different locations on the solar surface. Therefore, incorrect localisation of the HCS can drive large variations in the back-mapping of source regions for the solar wind  \citep{badman2023prediction}. Calibrating the coronal models with white light coronagraph observations of streamers and pseudo-streamers has somewhat improved the prediction of the HCS \citep{poirier2021exploiting, badman2022constraining}. However, each scan of the corona takes half a solar rotation to complete in which time the coronal field can evolve significantly \citep{edwards2022solar}. 

In this letter, we show that nested active regions have the potential to restrict the location of the HCS for multiple rotations, which could be used to improve the accuracy of spacecraft connectivity predictions. We first identify nested flux emergence from epochs of stalling in the reversal of the Sun's dipole magnetic field. Then we demonstrate how the nesting of active regions influences the coronal magnetic field and discuss how the resulting structure could be used to more accurately connect remote-sensing with in-situ observations.

\section{Identifying stalling epochs}

We search for time periods when the evolution of the Sun's coronal magnetic field is restricted due to nested active region emergence. To do this, we examine the dipole component of the Sun's photospheric magnetic field; as \citet{finley2023evolution} showed this can be influenced by strong flux emergence events. The reversal of the Sun's dipole magnetic field during the solar cycle is well documented, as well as its correlation with the higher-order magnetic field components \citep{derosa2012solar}. Magnetic field observations from both the Michelson Doppler Imager (MDI), on board the Solar and Heliospheric Observatory (SOHO), and the Helioseismic and Magnetic Imager (HMI), on board the Solar Dynamics Observatory (SDO) are used. We follow the methodology of \citet{finley2023evolution} by performing a spherical harmonic decomposition of the photospheric magnetic field using Carrington magnetograms from SOHO/MDI \citep{scherrer1995solar} and SDO/HMI \citep{scherrer2012helioseismic}. Both of these magnetogram time-series include a polar field correction \citep{sun2011new, sun2018polar}. From their decomposition, we reconstruct the dipole component of the magnetic field and extract the inclination of the positive dipole axis with respect to the rotation axis. The latitude of the positive pole is shown in Figure \ref{fig:br_lat} along with the monthly sunspot number, taken from the World Data Center SILSO at the Royal Observatory of Belgium\footnote{Data accessed June 2024: https://www.sidc.be/SILSO/datafiles}. A 13-month box-car smoothing of each time-series is over-plotted with a dashed red line. Cycle to cycle variation in solar activity and the dipole reversal for sunspot cycles 23, 24 and 25 are discussed in Appendix \ref{ap:cycles}.

From the evolution of the dipole latitude in Figure \ref{fig:br_lat}, we find that each sunspot cycle contains multiple epochs where the reversal of the dipole magnetic field stalls. Periods of stalling that persist for six months to a year are manually identified and highlighted with horizontal bars. We also add horizontal bars to the monthly sunspot number above to show that these epochs of stalling are frequently associated with bursts of solar activity. Further investigation reveals that each of these time periods contains one or more region with nested flux emergence. This is summarised in Appendix \ref{ap:examples}. Three stalling epochs, from different sunspot cycles and phases of solar activity, are highlighted in magenta a), b), and c), and will be used to demonstrate the general behaviour of these nested active regions on the coronal magnetic field. From Figure \ref{fig:br_lat}, stalling epochs appear to be less frequent in sunspot cycle 23 than 24. This could be due to differences in the input magnetograms moving from SOHO/MDI to SDO/HMI or, more likely, is an indication that stalling epochs are easier to identify when the Sun's activity is weaker. This suggests that the influence of nested active regions is stronger during weaker sunspot cycles. 

\section{Results}

\begin{figure*}[h!]
    \centering
    \includegraphics[trim=0cm 0cm 0cm 0cm, clip, width=\textwidth]{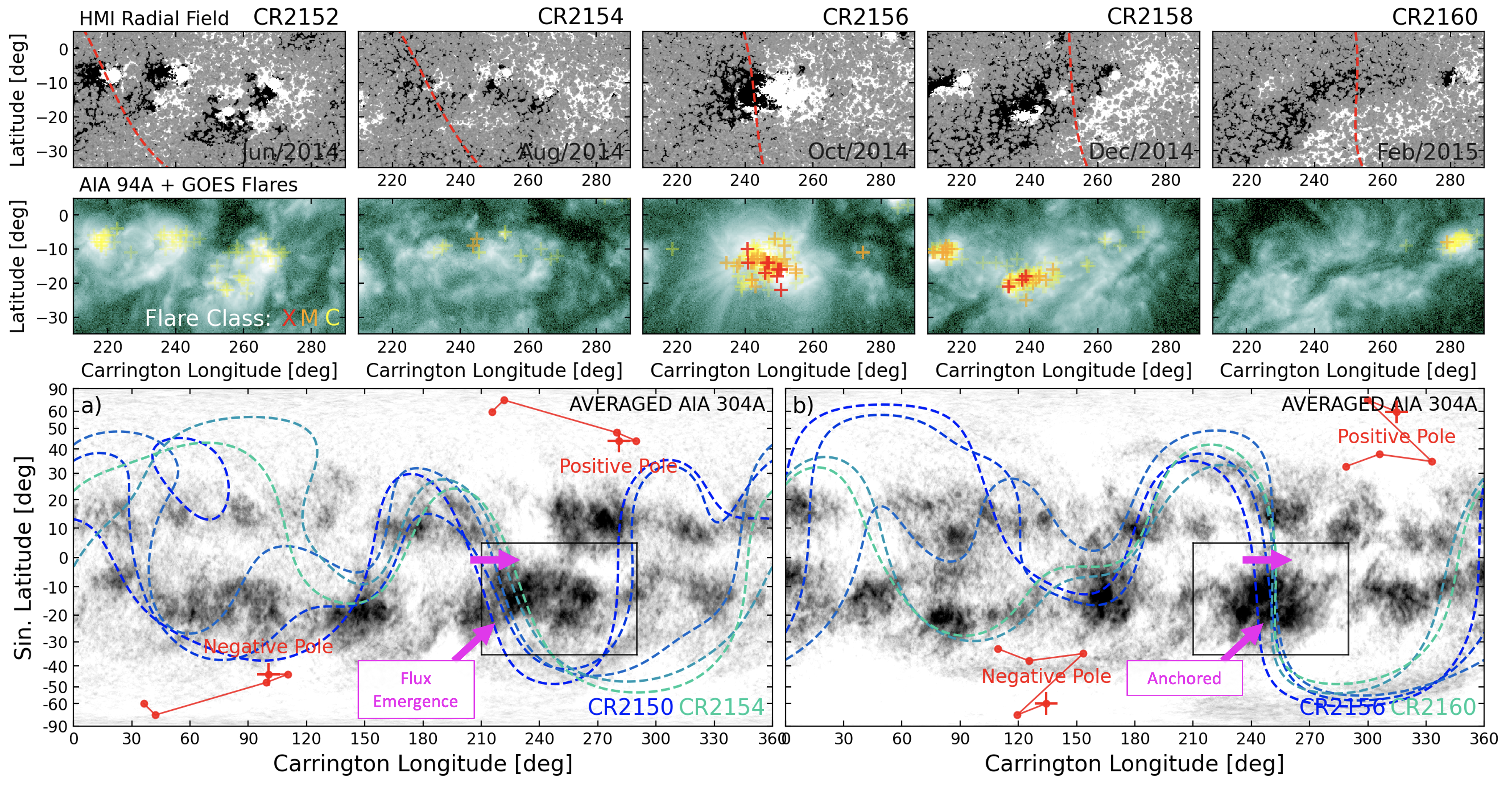}
    \caption{Same as Figure \ref{fig:fig_a}, but for nested active region emergence spanning June 2014 to March 2015. Flux emergence began in this region as early as 2013. The epoch highlighted here is slight later in spring 2014, when the region reached its peak strength and held the HCS in place for several solar rotations.}
    \label{fig:fig_b}
\end{figure*}

\begin{figure*}[h!]
    \centering
    \includegraphics[trim=0cm 0cm 0cm 0cm, clip, width=\textwidth]{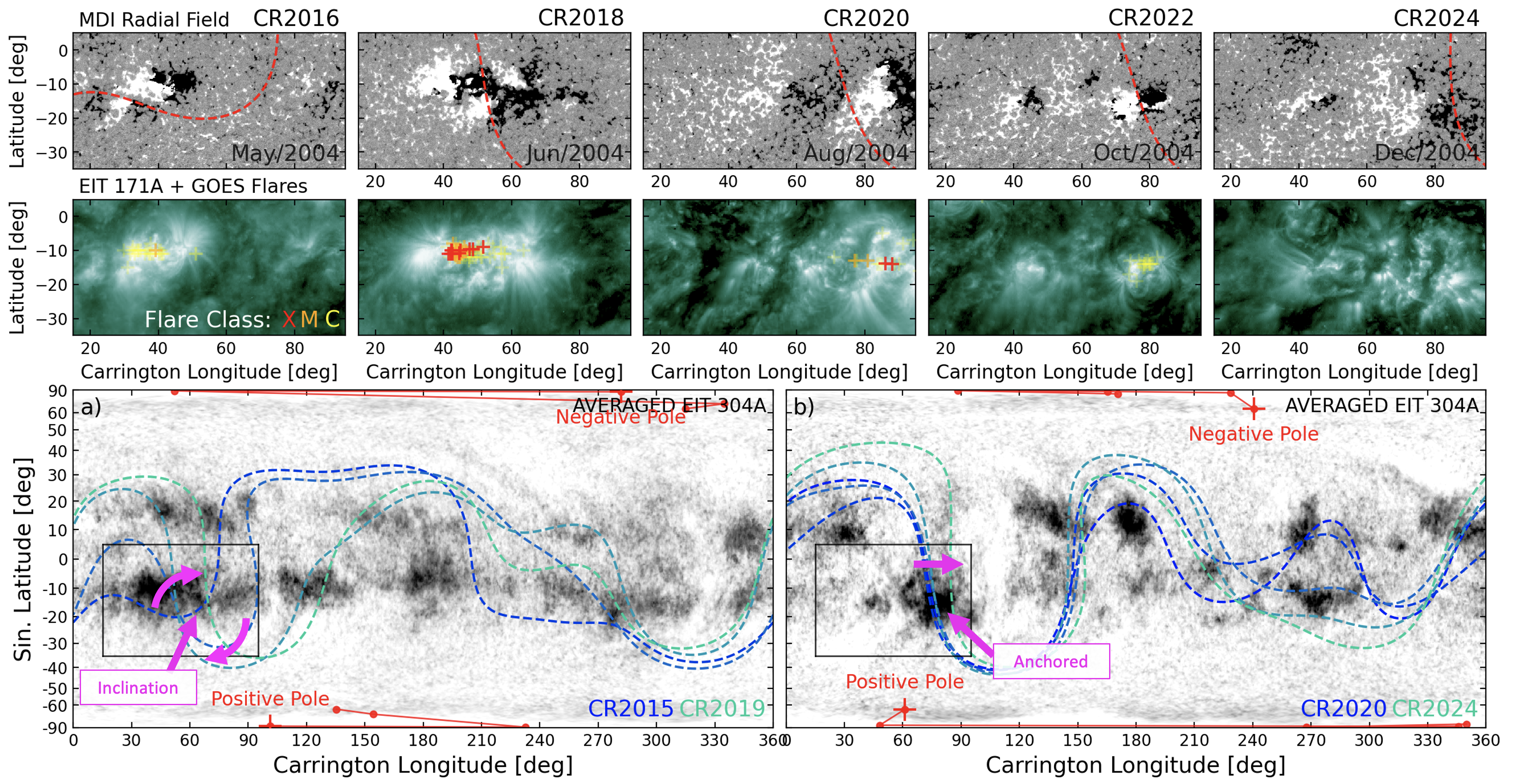}
    \caption{Same as Figure \ref{fig:fig_a}, but for a nested active region emergence spanning May 2004 to April 2005. The nested active region locally inclines the heliospheric current sheet, which then remains anchored to the region as it begins to drift with the rotation near the equator.}
    \label{fig:fig_c}
\end{figure*}

We explore the connection between the photospheric and coronal magnetic field for three of the stalling epochs identified in Figure \ref{fig:br_lat}; each containing a nested active region. We follow changes in the photospheric magnetic field, extreme ultra-violet (EUV) emission, and occurrence rate of x-ray flares, in addition to assessing the evolution of the coronal magnetic field and the HCS. The latter involves the extrapolation of the coronal magnetic field from photospheric magnetograms using a PFSS model. The equations and methodology used for this are fully described in \citet{finley2023accounting}. In this study, the source surface radius is fixed at 2.5 solar radii.

The PFSS model assumes a force-free coronal magnetic field with no currents, which cannot reproduce the twisted small-scale magnetic field of the emerging active regions. However, the emphasis of this study is on the large-scale structure of the corona for which the PFSS provides a good first approximation \citep{riley2006comparison}. We examine the rotation to rotation evolution using Carrington magnetograms, with each Carrington rotation (CR) having a duration of 27.28 days. However, in future, observations from Solar Orbiter will be combined to produce a more continuous set of observations (Finley et al. in prep).

\subsection{a) stalling epoch of 2022 (rising phase)}

We first examine one of the more recent stalling epochs from 2022, labelled with a magenta a) in Figure \ref{fig:br_lat}. During this time, sunspot cycle 25 was in the rising phase and magnetic activity was evenly distributed between the northern and southern hemispheres. The nested active region responsible for the stalling is located in the northern hemisphere around 20$^{\circ}$ latitude and migrated due to differential rotation over several solar rotations from 100$^{\circ}$ Carrington longitude to 40$^{\circ}$. Activity in this region first began on the Sun's far-side, detected with helioseismic inversions \citep{lindsey2017seismic}. Once this region rotated into view of Earth in January 2022, it remained a persistent feature during 2022. 

Figure \ref{fig:fig_a} shows the evolution of this region from CR 2254 to 2262; spanning Feburary 2022 to October 2022. The top row shows the photospheric magnetic field from SDO/HMI with two CRs between each column. Changes in the magnetic field configuration correspond to the appearance of new active regions and the diffusion of previously emerged magnetic flux (isolated active regions typically decay after one or two CRs). This nested active region reached its largest unsigned magnetic flux during April/May 2022. The second row shows the 94$\AA$ emission from SDO/AIA along with markers indicating the location of flares observed by GOES (each coloured by flare class). From the EUV emission and frequency of x-ray flares, activity in this region peaked accordingly during April/May 2022. The bottom two panels, a) and b), split the time period of interest into two with each containing information spanning five CRs. The background grey-scale represents the averaged SDO/AIA 304$\AA$ emission during those five respective CRs over the entire Sun, which is useful for highlighting persistently active features. A black box identifies the nested active region under investigation. The motion of the positive and negative pole of the dipole component are shown with connected red dots, with the crossed marker being the last in the sequence of magnetograms. The evolution of the HCS from the PFSS modelling is shown with coloured dashed lines. Each line represents the HCS for a given CR with the colour advancing from blue to green.

During the first five CRs, shown in panel a), this region grew in strength with multiple flux emergence events. This lead to increased flaring activity from this region (some X-class). As the solar cycle was just beginning, this nested active region produced a significant fraction ($\sim 40\%$) of the total solar flares during 2022 (Finley et al. in prep). This fraction is even higher if we consider flares from just the northern hemisphere. As the region grew in strength, the initially flat equatorial HCS from solar minimum became inclined, with the nested active region acting as a pivot. The strong oppositely directed magnetic field forms large closed magnetic field loops that anchored the HCS above. This can be seen more clearly in the top row of Figure \ref{fig:fig_a} where the HCS is plotted with a red dashed line over the photospheric magnetic field. Both Parker Solar Probe (encounter 11) and Solar Orbiter passed through this region during CR 2254.

During the following five CRs, shown in panel b), the active regions began to decay and the HCS drifts in longitude. This is due to the surface differential rotation acting upon the region. So despite containing enough magnetic flux to anchor the HCS, the region was no longer strong enough to resist the surface flows. At the same time, the dipole axis is observed to be advected in longitude with this region. Ultimately, a large filament forms between the oppositely directed polarity field, visible during CR 2260 in Figure \ref{fig:fig_a}. The behaviour of this region appears typical of the nested active regions identified from Figure \ref{fig:br_lat}, i.e. those that stall the reversal of the dipole magnetic field. In general, nested active region emergences modify the HCS and then, once the HCS is anchored, govern its relative position following the surface flows or subsequent flux emergence events. We will explore two more examples of this behaviour.

\begin{figure*}
    \centering
    \includegraphics[trim=0cm 0cm 0cm 0cm, clip, width=\textwidth]{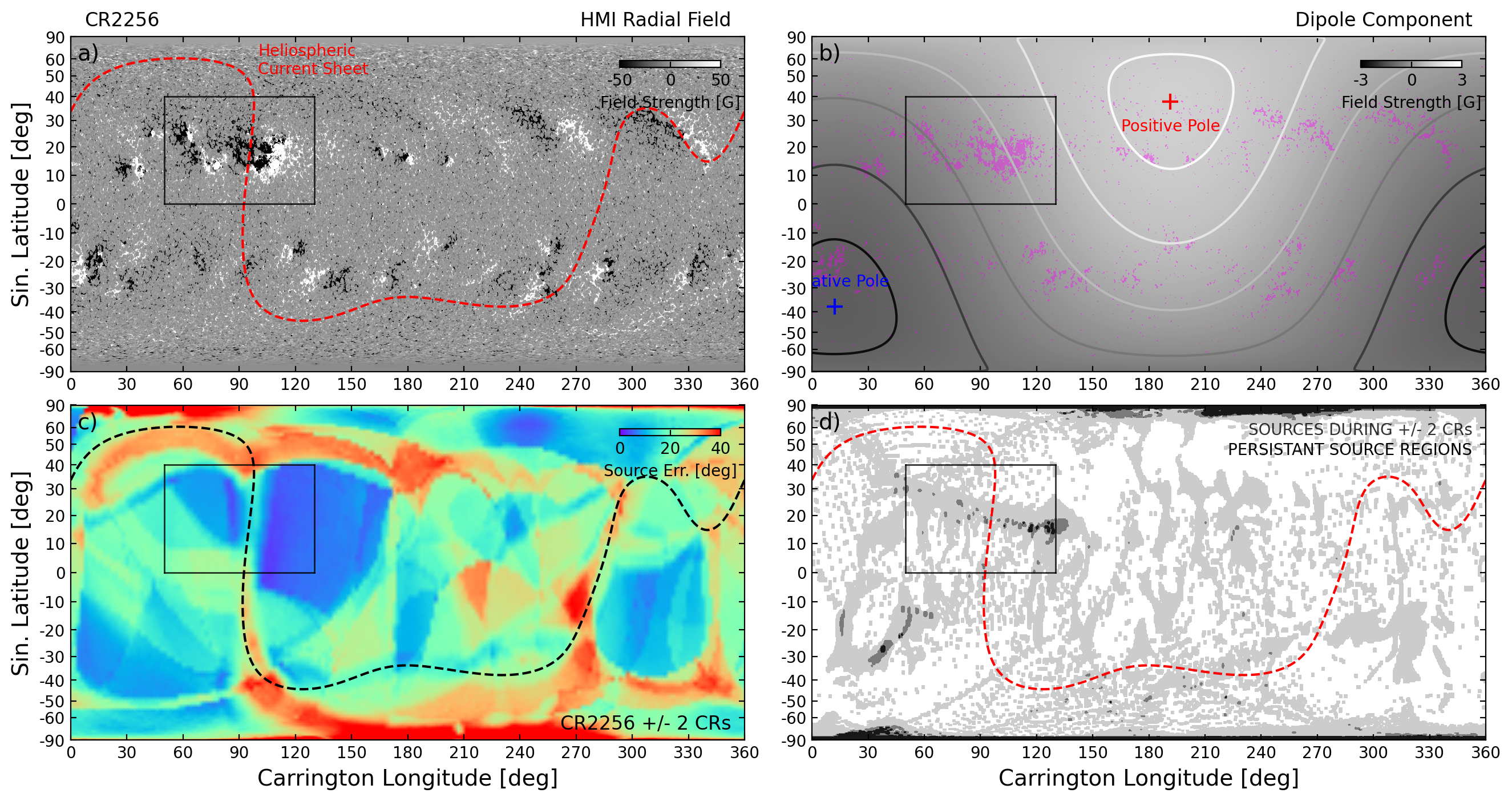}
    \caption{Summary of the solar wind connectivity for CR 2256 with the nested active region from Figure \ref{fig:fig_a}. Panel a) shows the photospheric magnetic field with the heliospheric current sheet in red and a box highlighting the area of interest. Panel b) shows the dipole component of the photospheric magnetic field from the spherical harmonic decomposition, along with the location of the dipole axis, and strong field regions from panel a) in magenta. Panel c) shows the variation in source location when back-mapping through PFSS models between five magnetograms spanning two CRs before and two CRs after CR 2256. Panel d) shows all the solar wind sources during the five consecutive CRs in grey, and persistent source regions across all CRs in black.}
    \label{fig:connect}
\end{figure*}

\subsection{b) stalling epoch of 2014 (activity maximum)}

The second stalling epoch selected from Figure \ref{fig:br_lat} spans from 2014 to 2015, labelled with a magenta b). During this time, the maximum of sunspot cycle 24 was ending and the declining phase of magnetic activity was beginning. The southern hemisphere was much more active than the northern hemisphere. The nested active region is located in the southern hemisphere, at around -15$^{\circ}$ latitude and migrated due to multiple flux emergence events as well as surface flows from around 210$^{\circ}$ Carrington longitude to 260$^{\circ}$ over several rotations. Activity was present in this region from as early as 2013, and declined during mid-2015.  

Figure \ref{fig:fig_b} summaries one of the most active time periods for this nested active region. The HCS is anchored to the magnetic field from this region in CR 2152 (June 2014). The averaged 304$\AA$ emission in panel a) shows this region to be a hot spot of activity, with persistent EUV emission. The emergence of new magnetic flux forces the HCS to migrate towards positive longitudes with respect to the Carrington frame from CR 2150 to 2154; spanning June to September 2014. This suggests that the underlying source of newly emerging flux is being steadily advected by the Sun's internal differential rotation. During this time, the dipole axis drifts in longitude with little variation in latitude. At CR 2156 in November 2014, the region reaches its peak unsigned flux, and no longer moves with respect to the Carrington frame. Panel b) shows the HCS was anchored in place for several solar rotations. The strength of the active region lead to increased EUV and flare activity (multiple X-class), after which the region steadily decayed. However, the HCS still remained anchored between the decaying polarities. A large filament forms during CR 2160, visible in Figure \ref{fig:fig_b}.

\subsection{c) stalling epoch of 2004 (declining phase)}

The third stalling epoch selected from Figure \ref{fig:br_lat}, labelled with a magenta c), took place in 2004. As this was before the launch of SDO, we use magnetic field and EUV observations from SOHO. This nested active region appeared in the declining phase of activity in sunspot cycle 23, during which time magnetic activity was evenly distributed between the northern and southern hemispheres. The nested active region was located in the southern hemisphere, at around -10$^{\circ}$ latitude and migrated due to latitudinal differential from around 30$^{\circ}$ Carrington longitude to 110$^{\circ}$ over several rotations. Activity was present in this region from January 2004, and declined during mid-2005.

This nested active region restricted the position of the HCS for several CRs, visible in Figure \ref{fig:fig_c}. Panel a) shows how the emergence of active regions modified the coronal magnetic field; inclining and anchoring the HCS above. Peak unsigned flux and flare activity was reached during CR 2018 (July 2004), at which point the HCS was fully inclined between the strong opposite polarities present in the region. Panel b) shows the HCS anchored to the magnetic field from the nested active region and, under the influence of surface differential rotation, drifting towards positive Carrington longitudes. Here both new flux emergence and surface motions played a role in advecting the active regions and the HCS. The dipole axis is almost aligned with the rotation axis and drifts in longitude along with the nested active region. 

This nested active region has a similar behaviour to that of the first (from 2022). Both took place away from solar maximum and so the role of nested flux emergence may have been more significant. In each case, the repeated emergence of active regions significantly warps the HCS and then the region is steadily advected by the local latitudinal differential rotation. The first example took place during the rising phase of activity in 2022, and so the individual active regions emerged at higher latitudes (following the butterfly pattern) and were advected backwards in longitude with respect to the Carrington frame. The third example occurred during the declining phase, and was influenced by rotational flows nearer the equator moving the region towards positive longitudes. The second example occurred at the end of an activity maximum and so the distribution of active regions over the solar surface was more complex. Nevertheless, the active regions that emerged controlled the position of the HCS and drifted towards positive longitudes due to its low-latitude emergence.

\section{Discussion and conclusions}

In this letter, we outline how the nesting of active regions influences the topology of the coronal magnetic field and can effectively anchor the HCS in place for multiple solar rotations. The coherent large-scale magnetic field that is produced from repeated flux emergence events is visible in the behaviour of the Sun's dipole magnetic field axis. Both the stalling in latitude and the drift in longitude are driven by the emergence and advection of nested active regions. The initial emergence of activity warps the HCS, and further flux emergence events reinforce the closed field region that holds the HCS in place. As the local magnetic field strength weakens, via diffusion or cancellation, the region begins to follow the surface motions which shift the dipole axis and HCS. Due to the Sun's latitudinal differential rotation, the direction of this motion changes during the solar cycle. In the rising phase of the cycle, active regions emerge at higher latitudes and drift backwards in Carrington longitude but as the cycle progresses active regions begin to emerge closer to the equator and so drift towards positive longitudes. 

The contribution of the nested active region from 2022 (Figure \ref{fig:fig_a}) to the dipole component of the photospheric magnetic field is highlighted in Figure \ref{fig:connect}. Panel a) shows a Carrington magnetogram of this region at its peak unsigned magnetic flux (and flaring activity) during CR 2256 with the HCS indicated with a dashed red line. Panel b) shows the dipole component extracted from the magnetogram, with the strong field regions from panel a) over-plotted in magenta. The equator of the dipole component runs directly through the nested active region. This region is of a sufficient size and strength to contribute to the dipole component, and the coronal magnetic field forms a strong closed field region above. This configuration persists even as the region decays.

Figures \ref{fig:fig_a}, \ref{fig:fig_b}, and \ref{fig:fig_c}, each show examples of nested active region emergences that influence the reversal of the coronal magnetic field (see Figure \ref{fig:br_lat}). The control that nested active regions have on the location of the HCS can be seen in each example when comparing the variability of the HCS between CRs at other longitudes away from the nested active region. This is most evident in Figure \ref{fig:fig_b}, during activity maximum, as the HCS between Carrington longitudes $0^{\circ}$ to $180^{\circ}$ varies wildly in comparison to the HCS above the nested active region. For this reason, identifying nested flux emergence can be useful when planning remote-sensing observations for modern heliophysics mission, given that the connectivity around these regions may be more reliably predicted from one CR to another.

Figure \ref{fig:connect}, panels c) and d) explore the variation in connectivity of the coronal magnetic field between consecutive CRs. Panel c) quantifies how much the footpoint of a magnetic field line, traced down from from the source surface, changes over five CRs centered on CR 2256, i.e. for each CR 2254 to 2258 the field line is traced down and the variation in footpoint position is evaluated across all extrapolations. The largest variations are found near the HCS as this divides areas with very different source regions. However, examining the source surface above the nested active region, there is a strong suppression of this effect as field lines are traced down consistently from one CR to another. Panel d) shows the sources of the open magnetic field at the solar surface for any of the five CRs in grey, with persistent sources for all extrapolations in black. We may therefore be more confident in connecting the solar wind back to these persistent sources that are supported by nested flux emergence. \citet{rivera2024situ} recently performed a conjunction study using Parker Solar Probe and Solar Orbiter during CR 2254. As this nested active region was emerging during this time, it may have increased the likelihood of finding a period of coherent connectivity between the two observers.

By following the long-term evolution of the Sun's magnetic field, and carefully monitoring the emergence history of active regions, it may be possible in future to provide more reliable remote-sensing targets for missions such as Solar Orbiter. Given that a large fraction of active regions appear to be nested, and that this study was limited by manual selection, a more thorough investigation is warranted. The community would benefit from incorporating the emergence history of active regions when studying the magnetic connectivity of the solar wind.

\begin{acknowledgements}
This research has received funding from the European Research Council (ERC) under the European Union’s Horizon 2020 research and innovation programme (grant agreement No 810218 WHOLESUN), in addition to funding by the Centre National d'Etudes Spatiales (CNES) Solar Orbiter, and the Institut National des Sciences de l'Univers (INSU) via the Programme National Soleil-Terre (PNST).
Data supplied courtesy of the SDO/HMI and SDO/AIA consortia. SDO is the first mission to be launched for NASA's Living With a Star (LWS) Program.
SOHO is a project of international cooperation between ESA and NASA. 
The sunspot number used in this work are from WDC-SILSO, Royal Observatory of Belgium, Brussels. 
Data manipulation was performed using the numpy \citep{2020NumPy-Array}, scipy \citep{2020SciPy-NMeth}, and pySHTOOLS \citep{wieczorek2018shtools} python packages.
Figures in this work are produced using the python package matplotlib \citep{hunter2007matplotlib}.
\end{acknowledgements}

%
%

\bibliographystyle{yahapj}
\bibliography{adam}

\begin{appendix}

\section{Sunspot cycle variation in activity and dipole field reversal}\label{ap:cycles}

\begin{figure*}[h!]
    \centering
    \includegraphics[trim=0cm 0cm 0cm 0cm, clip, width=\textwidth]{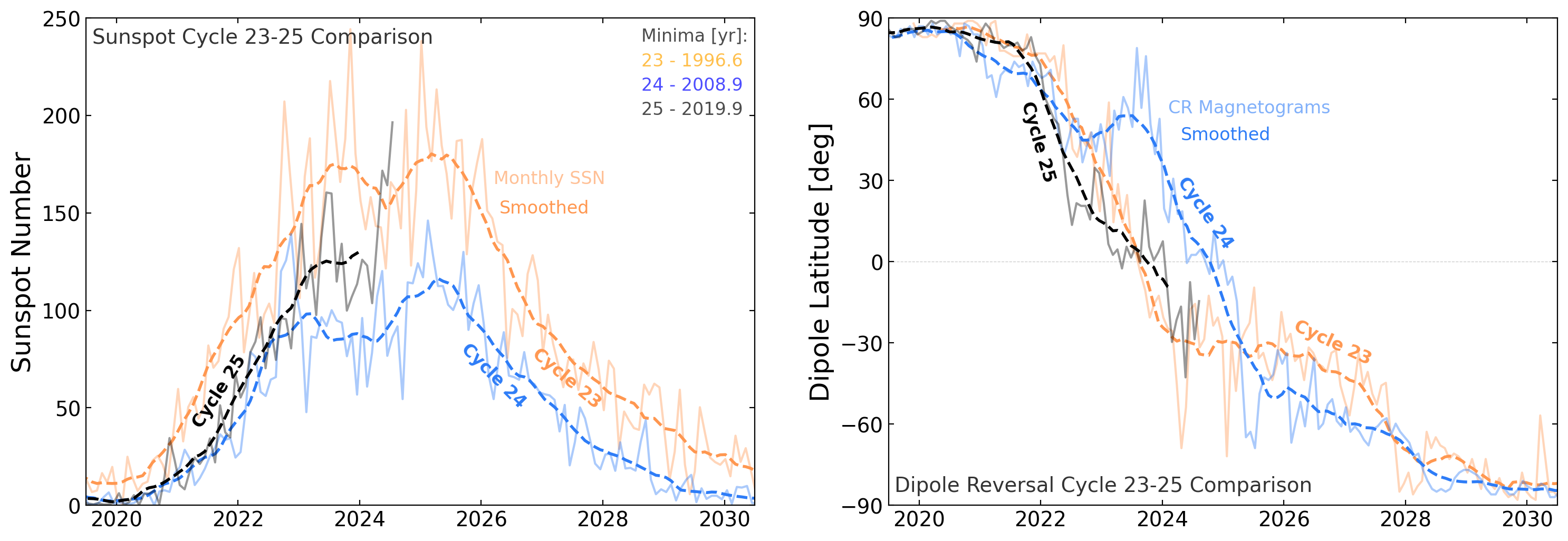}
    \caption{Comparison of the monthly sunspot number and positive dipole axis evolution between cycles 23, 24, and 25, shown in orange, blue, and black, respectively. The start times of cycles 23 and 24 have been shifted to match that of cycle 25. Dashed lines show the 13-month smoothed time-series.}
    \label{fig:stacked}
\end{figure*}

Each solar activity cycle has a duration of around 11 years that includes a sharp rising phase from activity minimum to maximum and then a longer declining phase from maximum back towards minimum. The maximum of activity in the northern and southern hemispheres are often shifted from one another by two to three years \citep{finley2023evolution}. As the total sunspot number grows, the frequency of energetic space weather events increases, from flares, to coronal mass ejections, and energetic particle events also rises \citep[see review of][]{temmer2021space}. The magnetic flux brought up to the surface in active regions subsequently decays and is advected by surface flows, leading to the reversal of the Sun's polar magnetic fields \citep{sun2015polar}. Figure \ref{fig:stacked} compares the monthly sunspot number for sunspot cycles 23, 24, and 25 (until August 2024), taken from the World Data Center SILSO at the Royal Observatory of Belgium, along with the reversal of the Sun's dipole magnetic field extracted from the spherical harmonic decomposition of photospheric magnetograms in Section 2.

Until recently, sunspot cycle 25 appeared to have a sunspot maximum in between that of cycles 23 and 24. However, a recent burst of activity from the southern hemisphere has pushed the monthly sunspot number up to the strength of cycle 23. The reversal of the dipole magnetic field shows some similarity between cycles 23 and 25. Typically, the reversal crosses the equator after three to fours years from the activity minimum. Cycle 24 deviated from this trend due to a burst of flux emergence in the northern hemisphere two years into the cycle. This significantly stalled the reversal for around one year. Likewise, strong magnetic activity during the second peak of cycle 23 held the dipole axis around $-30^{\circ}$ latitude for more than two years. Cycle 25 is currently undergoing a similar burst of activity with the dipole axis fixed around $-25^{\circ}$ latitude. It is unclear if the similarity between the reversals of cycles 23 and 25 will continue.

\section{Nested active regions from Figure \ref{fig:br_lat}}\label{ap:examples}

\begin{figure*}[h!]
    \centering
    \includegraphics[trim=0cm 0cm 0cm 0cm, clip, width=\textwidth]{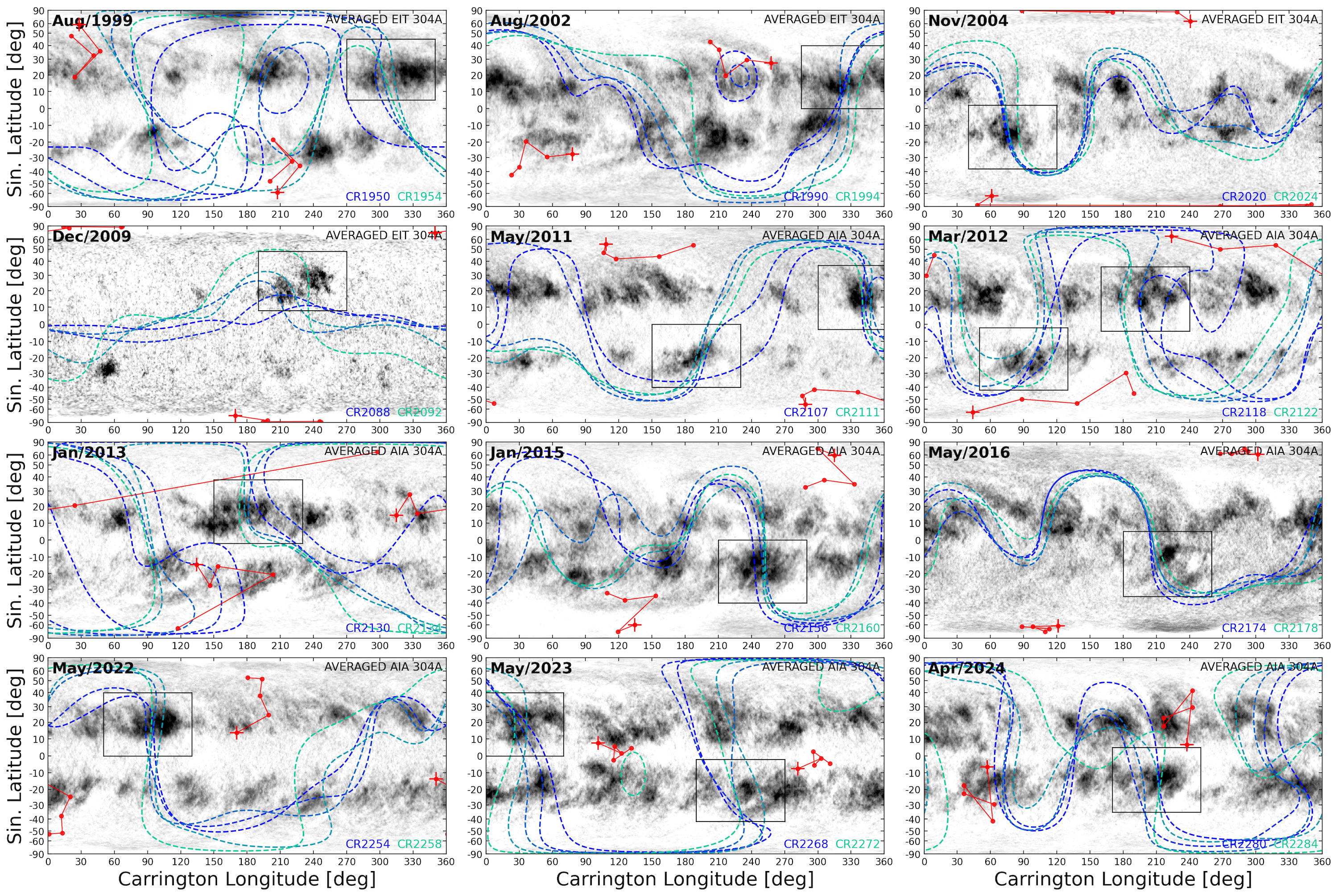}
    \caption{Identification of nested active regions. Each panel corresponds to one of the horizontal bars from Figure \ref{fig:br_lat}, corresponding to an epoch of stalling in the reversal of the Sun's dipole magnetic field. Data is combined from five CRs during each period, in a similar manner to Figures \ref{fig:fig_a}-\ref{fig:fig_c}, in order to highlight regions with persistent EUV emission (background grey-scale) that restrict the position of the HCS (dashed coloured lines) and motion of the dipole axis (red connected dots). Black boxes highlight regions with repeated flux emergence that anchor the HCS above. }
    \label{fig:more_examples}
\end{figure*}

For each of the epochs identified in Figure \ref{fig:br_lat} we search for regions with persistent activity by averaging the SDO/AIA 304$\AA$ emission over the entire Sun. The intensity of this is useful to visually identify hot spots for repeated flux emergence. The background grey-scale of each panel in Figure \ref{fig:more_examples} shows the EUV emission averaged over five CRs during each time interval. We present the most active five consecutive CRs from each stalling epoch, focusing on the influence that these hot spots of activity have on anchoring the HCS. During the same five CRs, the motion of the positive and negative axis of the dipole component of the Sun's photospheric magnetic field are shown with connected red dots. The evolution of the HCS from PFSS modelling is shown with coloured dashed lines. Each line represents the HCS for a given CR with the colour advancing from blue to green. This matches the analysis performed in panels a) and b) of Figures \ref{fig:fig_a}, \ref{fig:fig_b}, and \ref{fig:fig_c}. 

In each panel of Figure \ref{fig:more_examples}, black boxes highlight nested active regions that are influencing the position of the HCS. The regions frequently lie along the equator of the dipole component i.e. separating the positive and negative poles. The dipole component of the magnetic field appears to pivot around these regions, as in Figure \ref{fig:fig_a} panel a), or be advected in longitude as in Figure \ref{fig:fig_a} panel b). The longitudinal motion of the dipole, and drift of the HCS, is solar cycle dependent, with regions from rising phases drifting to the left (or remaining stationary) and maxima/declining phases drifting to the right. This is due to the influence of the latitudinal differential rotation on the nested active regions. The persistence of some nesting locations between panels (which can span one to three years) supports the claim that the underlying mechanism has a continuity on longer timescales than any single nested active region emergence \citep{berdyugina2003active}.

\end{appendix}

\end{document}